\documentclass[aps,prb,superscriptaddress,twocolumn,showpacs,amsmath,amssymb]{revtex4}
\usepackage{amsmath}
\usepackage{graphicx,epsfig}
\usepackage{bm}

\renewcommand{\v}[1]{{\bf #1}}
\newcommand{\be}{\begin{equation}}
\newcommand{\ee}{\end{equation}}
\newcommand{\bd}{\begin{displaymath}}
\newcommand{\ed}{\end{displaymath}}
\newcommand{\ba}{\begin{eqnarray}}
\newcommand{\ea}{\end{eqnarray}}
\newcommand{\nn}{\nonumber \\}

\newcommand{\bpm}{\begin{pmatrix}}
\newcommand{\epm}{\end{pmatrix}}

\begin{document}

\title{Skyrmion Lattice in Two-Dimensional Chiral Magnet}

\author{Jung Hoon Han}
%\email[Electronic address:$~~$]{hanjh@skku.edu}
\affiliation{Department of Physics, BK21 Physics Research Division,
Sungkyunkwan University, Suwon 440-746, Korea}
\author{Jiadong Zang}
\affiliation{Department of Physics, Fudan University, Shanghai
200433, China}\affiliation{Department of Applied Physics, University
of Tokyo, 7-3-1, Hongo, Bunkyo-ku, Tokyo 113-8656, Japan}
\author{Zhihua Yang}
\affiliation{Department of Physics, BK21 Physics Research Division,
Sungkyunkwan University, Suwon 440-746, Korea}
\author{Jin-Hong Park}
\affiliation{Department of Physics, BK21 Physics Research Division,
Sungkyunkwan University, Suwon 440-746, Korea}
\author{Naoto Nagaosa}
\affiliation{Department of Applied Physics, University of Tokyo,
7-3-1, Hongo, Bunkyo-ku, Tokyo 113-8656, Japan} \affiliation{
Cross-Correlated Materials Research Group (CMRG),
and Correlated Electron Research Group (CERG),
RIKEN-ASI, Wako, Saitama 351-0198, Japan}

\date{\today}

\begin{abstract} We develop a theory of the magnetic field-induced
formation of Skyrmion crystal state in chiral magnets in two spatial
dimensions, motivated by the recent discovery of the Skyrmionic phase
of magnetization in thin film of Fe$_{0.5}$Co$_{0.5}$Si and in the
A-phase of MnSi. Ginzburg-Landau functional of the chiral magnet
re-written in the CP$^1$ representation is shown to be a convenient
framework for the analysis of the Skyrmion states. Phase diagram of
the model at zero temperature gives a sequence of ground states,
helical spin $\rightarrow$ Skyrme crystal $\rightarrow$ ferromagnet,
as the external field $B$ increases, in good accord with the
thin-film experiment. In close analogy with Abrikosov's derivation of
the vortex lattice solution in type-II superconductor, the CP$^1$
mean-field equation is solved and shown to reproduce the Skyrmion
crystal state.
\end{abstract}

\maketitle

\section{Introduction}

Skyrmions, originally proposed as a model for baryons in nuclear
physics\cite{skyrme}, were first realized experimentally in the
condensed-matter system of quantum Hall ferromagnets near integer
filling factor $\nu \approx 1$\cite{sondhi,skyrmion-evidence}. The
electrons spontaneously form a fully polarized ferromagnet at $\nu =
1$ due to the exchange interaction, while slightly away from it the
spins of extra (or lack thereof) electrons organize themselves into
an intricate Skyrmionic structure as a result of the competitive
interplay between the Zeeman and the Coulomb
interactions\cite{sondhi}. It was further suggested that these
Skyrmions may condense into a crystalline form\cite{macdonald}.
Skyrmions as quasiparticle excitations in quantum Hall ferromagnets
are by now well established experimentally\cite{skyrmion-evidence}.
Recent experiments also find support for the crystallization of
Skyrmions in the quantum Hall system\cite{skyrme-crystal-evidence}.

More recently, a strong case for the formation of Skyrme crystal
(SkX) state was presented in the A-phase of a metallic ferromagnet
MnSi\cite{hexagonal-SC} and other compounds of B20
structure\cite{FeCoSi,B20-general}. Here, the main evidence comes
from small angle neutron diffraction data which exhibits a clear
hexagonal pattern consistent with the triangular lattice arrangement
of Skyrmions. Subsequent observation of the anomalous Hall effect in
the A-phase gives additional support to the Skyrmion lattice
picture\cite{AHE}. On the other hand, Monte Carlo simulation of the
two-dimensional classical spin model with the Dzyaloshinskii-Moriya
(DM) interaction and magnetic anisotropy predicted the existence of
various Skyrmion crystal phases as the ground state\cite{YONH}. Just
after this theoretical proposal, a real-space observation of the
Skyrmions forming a crystalline phase has been provided by the
Lorentz TEM imaging technique in the thin film of a
non-centrosymmetric magnetic crystal Fe$_{0.5}$Co$_{0.5}$Si under a
perpendicular magnetic field\cite{tokura}. By now there is
accumulating experimental evidence stating that Skyrmion lattice is a
natural occurrence in magnetic metals lacking inversion symmetry. For
such crystals the role of the DM interaction is to convert a $\v k
=\v 0$ ferromagnetic ground state to a helical phase at a finite
ordering vector, $\v k \neq \v 0$, with the  magnitude $|\v k|$ fixed
by the ratio of the DM exchange over the Heisenberg exchange
energies. Influence of thermal fluctuation and/or the magnetic field
can favor the formation of multiple spiral phase over a single spiral
(In this paper we use the words ``spiral" and ``helical"
interchangeably). When imaged in real space, the multiple spiral
phase is none other than the Skyrmion
lattice\cite{hexagonal-SC,tokura,YONH}.

We should emphasize the contrasting behavior of the chiral magnet
realized in two (2D) and three (3D) spatial dimensions. For 3D
systems such as examined in Refs.
\onlinecite{hexagonal-SC,FeCoSi,B20-general}, the Skyrmion crystal
forms over a small window of finite temperature just below the
paramagnetic transition known as the A-phase. On the other hand, for
quasi-2D chiral magnets recently synthesized with the thickness
smaller than a single spiral period\cite{tokura}, the Skyrmion phase
occurs over a much larger temperature range extending nearly up to
the paramagnetic transition  and down to the lowest temperature
measured, $T\approx$ 5K, strongly suggesting that the Skyrme crystal
is the ground state. Free energy analysis of the 3D model indeed
shows stability of the Skyrmion phase at a finite temperature, but
not at zero temperature\cite{hexagonal-SC,FeCoSi,B20-general}. As
shown in this paper, the 2D case supports the Skyrmion phase even at
zero temperature.

Bogdanov and collaborators had earlier investigated the instability
of the helical spin state to the spontaneous creation of Skyrmionic
spin texture due to magnetic field in chiral
magnets\cite{bogdanov1,bogdanov2}. Although not explicitly emphasized
at that time, present experimental
situation\cite{hexagonal-SC,FeCoSi,B20-general,tokura} shows that the
dimensionality plays a great role in stabilizing Skyrme crystal phase
at low temperature. Our paper therefore re-visits the issue of the
phase diagram of the chiral magnet under the influence of magnetic
field first pioneered by Bogdanov \textit{et al}, in light of these
recent developments. We base our calculation explicitly on two
dimensions, assuming spin models with zero spin anisotropy such as
known to be the case for Fe$_{0.5}$Co$_{0.5}$Si. As a theoretical
advance, we show that the Ginzburg-Landau action for the chiral
magnet can be cast in a simple form using the CP$^1$ representation
of the classical spin. The advantage offered by the new
representation becomes evident when we consider the lattice case,
where it becomes possible to derive the solution for the Skyrmion
lattice state by exploiting the analogy to another celebrated example
of the lattice of topological defects - Abrikosov vortex lattice.

This paper is organized as follows. In Sec. \ref{sec:CP1} we
introduce CP$^1$ formulation of the Ginzburg-Landau (GL) energy
functional for a chiral magnet. In Sec. \ref{sec:phase-diagram} we
derive the zero-temperature phase diagram of the 2D chiral GL model
using two complementary approaches. Both lead to the ground state
evolution with the external field in excellent accord with the recent
experiment. Analytical solution of the Skyrme lattice state is
presented. Finally we conclude with a brief, qualitative discussion
of thermal fluctuation effects for the Skyrmion crystal and summarize
in Sec. \ref{sec:discussion}. One can read this paper as a companion
article to Ref. \onlinecite{tokura}, where the experimental findings
and Monte Carlo results were reported.

\section{CP$^1$ Formulation of Chiral Ginzburg-Landau Theory}
\label{sec:CP1}

We are concerned in this paper with magnetic systems lacking an
inversion symmetry such as MnSi and Fe$_{1-x}$Co$_{x}$Si, where the
magnetic interactions are characterized by a ferromagnetic exchange
$J$ and a weaker, Dzyaloshinskii-Moriya coupling $D$. Hereafter, we
take the length of a structural unit cell as unity. The competition
of the two interactions leads to a helical spin ground state with the
pitch vector of length $k = D/J$. The effective continuum theory for
such chiral magnet is provided by the Ginzburg-Landau (GL) energy
functional\cite{bogdanov1,bogdanov2,bak}

\ba {\cal F}[\v n ] \!=\! {J\over 2} \sum_{\mu} (\partial_\mu \v n)
\cdot (\partial_\mu \v n) \!+\! D \v n \cdot \bm \nabla \times \v n
\!-\! \v B \cdot \v n . \label{eq:energy-density}\ea
The non-linear $\sigma$-model given in the first part describes the
excitation of a conventional ferromagnet while the second term gives
the effect due to the DM exchange. As we are interested in the
effectively two-dimensional chiral magnet such as recently
synthesized with Fe$_{0.5}$Co$_{0.5}$Si\cite{tokura}, the spatial
derivatives run over the two-dimensional plane, $\mu = x,y$. The
total energy is the spatial integral of $\cal F$: $F[\v n ]=\int d^2
\v r ~ {\cal F}[\v n ] $. It is also known that spin anisotropy does
not play an important role in Fe$_{0.5}$Co$_{0.5}$Si, which is
therefore omitted from the theory.

According to the Monte Carlo simulation\cite{YONH} and also the
recent experiment\cite{tokura}, the spiral spin states at $\v B =0$
is supplanted by the hexagonal packing of Skyrmions when $\v B$
oriented perpendicular to the thin film plane exceeds a certain
strength, $B_{c1}$. At still higher field strength $B>B_{c2}$ the
Skyrme crystal state gives way to fully polarized ferromagnet. Both
transitions are first-order, as evidenced by the presence of the
co-existence region around both $B_{c1}$ and $B_{c2}$\cite{tokura}.
For the intermediate phase $B_{c1}<B<B_{c2}$ the triangular lattice
of Skyrmions identified by the Lorentz TEM bears striking
resemblance to the Abrikosov vortex array in type-II
superconductors\cite{abrikosov,tonomura}. In this paper we correctly
reproduce the observed first-order transitions of magnetic phases
under increasing magnetic field, by comparing the energies of the
three candidate states: H (helical), SkX (Skyrme crystal), and FM
(ferromagnetic).

Previous theoretical
approaches\cite{bogdanov1,bogdanov2,bak,hexagonal-SC} were based on
the O(3) representation of the spin vector, $\v n = (\sin\theta \cos
\phi, \sin\theta \sin \phi, \cos\theta)$. In this work, we adopt
instead a ``complex" description in which $\v n$ is replaced by a
pair of complex fields

\ba\v z = \bpm z_1 \\
z_2 \epm = \bpm e^{-i\phi} \cos(\theta/2) \\
\sin (\theta/2) \epm,  \ea obeying the unit-modulus constraint $\v
z^\dag \v z = 1$\cite{auerbach}. Mapping to the spin vector is
through $\v n = \v z^\dag \bm \sigma \v z$. A well-known identity
allows the re-writing of the nonlinear sigma model in the CP$^1$
language (no sum on $\mu$)\cite{auerbach}

\ba & {1\over 4} (\partial_\mu \v n) \cdot (\partial_\mu \v n) =
(\partial_\mu \v z^\dag ) \cdot (\partial_\mu \v z)- A_\mu^2 , &\nn
& A_\mu = - {i \over 2}[{\v z}^\dag (\partial_\mu {\v z}) -
(\partial_\mu {\v z}^\dag ){\v z}] . &  \label{eq:CP1-NLsigmaM}\ea
The Skyrmion density is related to the magnetic field associated with
the vector potential by ${1\over 2} \v n \cdot (\partial_x \v n
\times
\partial_y \v n ) =  \partial_x A_y - \partial_y
A_x$. The chiral, DM term in the energy (\ref{eq:energy-density})
under the CP$^1$ mapping becomes

\ba \v n \cdot (\bm \nabla \!\times\! \v n ) \!=\! -2 \v n \cdot \v A
\!-\! i \v z^\dag (\bm \sigma \cdot \bm \nabla) \v z \!+\! i (\bm
\nabla\v z^\dag ) \cdot \bm \sigma \v z . \label{eq:CP1-DM}\ea
When Eqs. (\ref{eq:CP1-NLsigmaM}) and (\ref{eq:CP1-DM}) are combined,
the energy density re-written in the CP$^1$ representation takes on a
succinct form

\ba {\cal F}[ \v z] &=&  2J \sum_\mu  \Bigl( D_\mu \v z \Bigr)^\dag
\Bigl( D_\mu \v z \Bigr) -\v B \cdot \v z^\dag \bm \sigma \v z ,\nn
D_\mu &=&
\partial_\mu - iA_\mu - i \kappa \sigma_\mu.
\label{eq:energy-density-in-CP1}\ea
The derivative $D_\mu$ is a 2$\times$2 matrix due to the non-zero DM
exchange, $\kappa = D/2J$. This CP$^1$ version of the energy density
of a chiral magnet and the saddle-point equation which follows from
it form the basis of the subsequent analysis\cite{comment1}.

To orient the readers, we first discuss how to write down a
conventional magnetic state in the CP$^1$ language. The ferromagnetic
state for instance is written $\v z_\mathrm{FM} = \v z_0$, an
arbitrary constant spinor. The helical spin state - the ground state
of Eq. (\ref{eq:energy-density}) at zero magnetic field - is obtained
from FM by a position($\v r$)-dependent twist of the spin
orientation,

\ba \v z_\mathrm{H} = e^{i \kappa (\bm \sigma \cdot \hat{k}) (\v r
\cdot \hat{k})}\v z_0. \label{eq:SS-answer}\ea
The Pauli matrix $\bm\sigma$ appears above. It can be checked that
the associated spin configuration $\v n_\mathrm{H} = \v
z^\dag_\mathrm{H} \bm \sigma \v z_\mathrm{H}$ is indeed orthogonal to
the propagation vector direction $\hat{k}=\v k/|\v k|$ and rotates
with the pitch $k = 2\kappa = D/J$, provided the initial spin
orientation $\v n_0 = \v z_0^\dag \bm \sigma \v z_0$ is perpendicular
to $\hat{k}$. The vector potential $\v A$ for the helical spin state
is identically zero.

\begin{figure}[ht]
\includegraphics[scale=0.4]{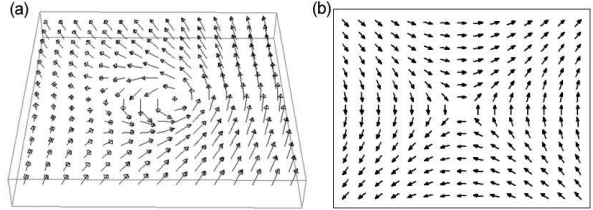}
\caption{(a) A typical (anti-)Skyrmion configuration given by Eq.
(\ref{eq:sample-Skyrmion}) with $R=2$. (b) $z_1$ component of the
CP$^1$ Skyrmion $\v z_\mathrm{Sk}$. When expressed as a planar spin
$(\mathrm{Re}[z_1], \mathrm{Im}[z_1])$, it is an anti-vortex.}
\label{fig:skyrmion-vortex-map}
\end{figure}

An isolated Skyrmion in two dimensions with the spin pointing down at
the origin $r=0$ and up far away, $r\rightarrow \infty$, is given the
CP$^1$ expression\cite{Raj}

\ba \v z_\mathrm{Sk} = e^{i(\theta(r) /2) (\bm \sigma \cdot
\hat{r}\times\hat{z})} \v z_0, \ea
where $\v z_0 = \bpm 1 \\ 0 \epm$, and $\hat{r} = (\cos \varphi, \sin
\varphi, 0)$ is the radial vector (We use $\phi$ for the azimuthal
angle of the spin vector $\v n$, and $\varphi$ for the azimuthal
angle of the coordinate $\v r = (x,y)$: $\tan \varphi = y/x$). The
polar angle $\theta (r)$ is a function of the radial coordinate $r$
such that $\theta(0)=\pi$, $\theta(\infty)=0$, and smoothly varying
in between. Vector potential for the single-Skyrmion configuration is
$\v A_\mathrm{Sk} = (\hat{\varphi}/r) \sin^2 (\theta /2)$. As a
concrete example of a single (anti-)Skyrmion one can take the
following case,

\ba n^z = {r^2 \!-\! R^2 \over r^2 \!+\! R^2} , ~~~ n^x \!+\! in^y =
2 i R {x\! +\! iy \over r^2 \!+\! R^2},
\label{eq:sample-Skyrmion}\ea
and the associated CP$^1$ expression $\v z_\mathrm{Sk} =
{1\over \sqrt{R^2 \!+\! r^2}} \bpm i x+ y  \\
-R \epm$. Note that the $z_1$ part of the CP$^1$ Skyrmion solution
behaves exactly as would a (anti-)vortex: depleted to zero at the
core, winding about the origin as $\sim e^{-i\varphi}$, and reaching
a constant unit magnitude when $r/R \gg 1$ (See Fig.
\ref{fig:skyrmion-vortex-map}). At least pictorially it is clear that
U(1) vortex configuration in the CP$^1$ representation (in $z_1$ for
instance) implies a Skyrmionic spin configuration in $\v n$ through
the mapping $\v n = \v z^\dag \bm \sigma \v z$. In the next section,
we will show that the mapping from a U(1) vortex to the real-spin
Skyrmion can be extended to the lattice case as well.

The spin configuration Eq. (\ref{eq:sample-Skyrmion}) is the saddle
point solution of the first, Heisenberg-only term of the free energy
given in Eq. (\ref{eq:energy-density}). It is an excited state, with
the energy $E=4\pi J$ independent of the radius of the Skyrmion $R$,
and carrying a non-trivial topological number\cite{Raj}. On the other
hand, the Skyrmion state can become the \textit{ground state} if we
consider the full free energy in Eq. (\ref{eq:energy-density}).
Although the exact spin configuration will be somewhat different from
that given in Eq. (\ref{eq:sample-Skyrmion}), the topological
properties remain the same. Furthermore, the radius $R$ is uniquely
determined by the ratio of DM interaction and the exchange energy as
shown in the next section. For Skyrmions in the quantum Hall system
the radius was fixed by the relative strengths of Zeeman and Coulomb
interactions\cite{sondhi}.

\section{Phase Diagram}
\label{sec:phase-diagram}

In this section we address the phase diagram of the model, Eq.
(\ref{eq:energy-density}) or Eq. (\ref{eq:energy-density-in-CP1}), as
the external field $\v B = B\hat{z}$, always assumed to be in the
positive $z$-direction ($B>0$), increases from zero. Experimentally,
the low-temperature phase evolves as helical spin, SkX, and FM with
increasing field strength. We adopt two approaches for calculation of
the energies of the respective phases, both of which successfully
reproduce the observed phase evolution. In addition, we obtain
analytic results for the two critical fields $B_{c1}$ and $B_{c2}$
($B_{c1}<B_{c2}$), each referring to the first-order critical field
separating H from SkX $(B_{c1})$, and SkX from FM $(B_{c2})$.

\subsection{Variational analysis of a single Skyrmion}
\label{subsection-a}

In this subsection, we regard the Skyrmion lattice as the
close-packing of individual Skyrmions of radius $R$ forming a
triangular lattice. The local spin orientation $(\theta, \phi)$ of a
single Skyrmion depends on the local coordinate $(r,\varphi)$ as
$\phi=\varphi-\pi/2$ and $\theta=\theta(r)$. The total free energy of
a single Skyrmion reads

\begin{eqnarray}\label{eq:Skyrmion}
F_\mathrm{Sk}&=&2J\int 2\pi rdr \Bigl[\left( \frac{1}{2}\frac{d\theta
}{dr}+\kappa \right)^{2}-
\kappa^{2}+\frac{\kappa}{%
r}\sin \theta \cos \theta \nn &&~~~~ +\frac{1}{4r^{2}}\sin ^{2}\theta
-\beta (\cos\theta-1) \Bigr],
\end{eqnarray}
where $\beta=B/(2J)$, and the FM state is chosen to have the free
energy zero. Ferromagnetic configuration is enforced in the outermost
region by the upward magnetic field, so that $\theta(\infty)=0$. On
the contrary, $\theta(0)=\pi$ due to the geometric nature of
Skyrmion. In numerical calculations, a hard radical cutoff $R$ is
introduced such that $\theta(r)=0$ for $r\geq R$. In physical terms
$R$ is half the inter-Skyrmion distance in the Skymre crystal. The
Skyrmion lattice observed experimentally can be constructed as the
close-packing of non-overlapping, individual Skyrmions in a
trianglular lattice. One can write down the total free energy of the
Skymre crystal state as

\ba F_\mathrm{SkX} =\frac{L^2}{2\sqrt{3}R^2}F_\mathrm{Sk},\ea
$L$ being the sample size. For each cutoff $R$, we can apply the
numerical variation to minimize the free energy $F_\mathrm{Sk}$ for a
single Skyrmion. The equilibrium configuration of the whole lattice
should minimize $F_\mathrm{SkX}$, hence the free energy density
functional $F_\mathrm{Sk}[\theta(r)]/R^2$ by optimizing $\theta(r)$.

%%%%%%%%%%%%%%%%%%%%%%%%%%%%%%%%%%%%%%%%%%%%%%%%%%%%%%%%%%%%%%%%%%%%%%%
\begin{figure}[tps]
\includegraphics[scale=0.4]{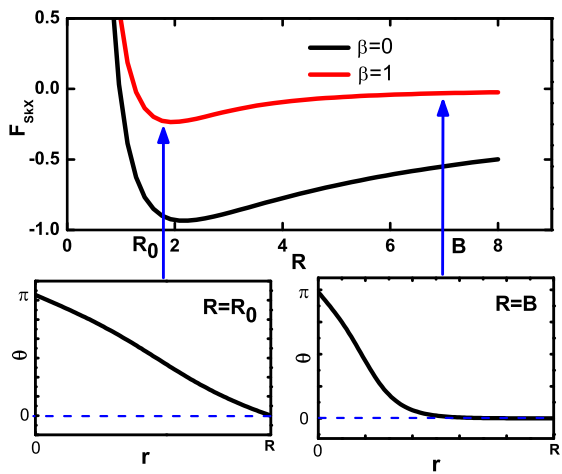}
\caption{(color online) Dependence of the free energy density
$F_\mathrm{SkX}$ on the radical cutoff $R$ at magnetic field
$\beta=1$ (red curve) and $\beta=0$ (black curve), respectively. We
choose $\kappa=1$. The angle $\theta$ measured from the $z$-axis as a
fuction of the distance $r$ from the center of the Skyrmion for two
different $R$'s are shown in the insets. Optimal $R_0$ for $\beta=1$
leads to an almost linear dependence of $\theta(r)$ on $r$, while
larger $R=B$ leads to a long ferromagnetic tail outside the Skyrmion
core region with radius $\cong R_0$. } \label{fig:density}
\end{figure}
%%%%%%%%%%%%%%%%%%%%%%%%%%%%%%%%%%%%%%%%%%%%%%%%%%%%%%%%%%%%%%%%%%%%%%%%%%%%%%%%%

Figure \ref{fig:density} shows dependence of the free energy
$F_\mathrm{SkX}$ on the cutoff $R$, where the sample size is
normalized. It clearly shows that when $R$ increases from zero,
$F_\mathrm{SkX}$ decreases dramatically, and reaches a minimal value
at $R_0$. This point is exactly the ground state of the helical
magnet at zero temperature where the compromise between Zeeman,
Heisenberg and DM interactions are reached to the maximum degree. The
configuration of $\theta(r)$ is shown in the left inset of Fig.
\ref{fig:density}. One can find that $\theta(r)$ varies almost
linearly in the whole region from $0$ to $R_0$, and reaches zero at
$R_0$. In this case, $R_0$ defines both the typical Skyrmion size as
well as half of the optimal inter-Skyrmion distance, and the
Skyrmions are close-packed. Applying the dimensional analysis to Eq.
(\ref{eq:Skyrmion}), one finds $r$ has the same dimension as
$1/\kappa\sim J/D$. Therefore it is expected that $R_0$ would be
proportional to $1/\kappa$, which is confirmed numerically as well.

When the cutoff $R$ exceeds the optimal distance $R_0$, the free
energy starts to increase from the negative minimal value and
approaches zero as $R\rightarrow \infty$. From the corresponding
configuration shown in the right inset of Fig. \ref{fig:density} one
can see that at $R>R_0$, the variation of $\theta(r)$ is
qualitatively different from that at $R=R_0$. A ferromagnetic tail
with $\theta=0$ appears over the region between $R_0$ and $R$. In
usual conventions, only the core region where the variation in
$\theta(r)$ is nonzero is referred as a Skyrmion. Therefore two
neighboring Skyrmions are well-separated by the intervening FM phase
in this case, and the Skyrmions are not close-packed.
%Some signature of the diluted Skyrmion packing was found in the Monte
%Carlo simulation near the SkX/FM boundary\cite{tokura,YONH}. In
%general, our analysis shows that the Skyrmion radius $R_0$ need not
%equal the inter-Skyrmion distance $R$.

The other competing phase observed besides the Skyrmion lattice is
the spiral configuration at small magnetic field.  This helical phase
is the exact ground state of the Hamiltonian containing the DM
interaction when the external field is absent. To see how the free
energy of the helical state varies with field, consider the spiral
spin propagating along $y$ direction with $\phi =0$ and
$\theta=\theta(y)$ is a function of $y$. The free energy for half
period of spiral is
\begin{equation}\label{eq:spiral}
F_\mathrm{H}\!=\!\frac{2JL}{\ell}\int_0^\ell dy
\left[\left(\frac{1}{2}\frac{d\theta
}{dy}\!+\!\kappa\right)^{2}\!-\!\kappa^{2}\!-\!\beta
(\cos\theta\!-\!1) \right],
\end{equation}
with the boundary condition $\theta(0)=\pi$ and $\theta(\ell)=0$,
where $2\ell$ is the period of a spiral. Similar to the Skyrmion
lattice case, one can derive the optimal $\ell$ by minimizing this
free energy functional to get the configuration for a spiral lattice.

\begin{figure}[tps]
\includegraphics[scale=0.4]{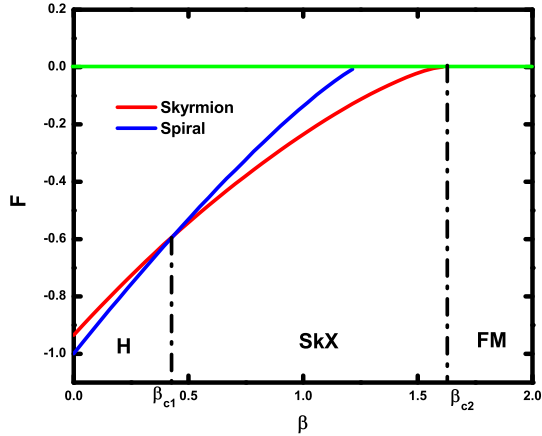}
\caption{(color online) The energies of the three states, i.e., (i)
helical state, (ii) Skyrmion crystal, and (iii) ferromagnetic state.
The free energy of ferromagnetic order is set to be zero, which is
labeled in green. The helical state is energetically favored at low
magnetic field, while the Skyrmion phase emerges at intermediate
magnetic field. Ferromagnetism is favored at larger fields.  }
\label{fig:phase}
\end{figure}

The free energies of the three phases (helical, SkX, FM) for
$\kappa=1$ obtained by variation are shown in Fig. \ref{fig:phase}.
It is explicitly shown that the spiral state has the lowest energy
among the three configurations when the external magnetic field is
small. In the large field limit, ferromagnetism is energetically
favored. In the intermediate region, one finds the Skyrmion phase as
a compromise between the DM and Zeeman energies. This is because the
Skyrmion configuration keeps the spiral structure inside and
ferromagnetic state outside, and hence can gain both the DM energy
and magnetic field energy. This result is quite consistent with the
Monte Carlo simulation\cite{YONH} and the experiment\cite{tokura}.
The lower ($\beta_{c1}$) and upper ($\beta_{c2}$) critical field
strengths separating the Skyrmion phase from other phases can be
easily determined by the intersects of Skyrmion energy line with the
other two in Fig. \ref{fig:phase}. Applying dimensional analysis to
Eq. (\ref{eq:Skyrmion}) or Eq. (\ref{eq:spiral}), the normalized
magnetic field $\beta$ is found to have the same dimension as
$\kappa^2$. Therefore one expects both $\beta_{c1}$ and $\beta_{c2}$
to scale with $\kappa^2$, a fact also established by numerical
calculation. Restoring proper units, we get the two critical fields

\ba B_{c1}=0.2D^2/J, ~~~ B_{c2}=0.8D^2/J. \label{eq:Bc1-Bc2} \ea

For Fe$_{0.5}$Co$_{0.5}$Si thin film, the two critical fields at low
temperatures were $B_{c1} \approx 40$mT and $B_{c2} \approx
80$mT\cite{tokura}. Due to the co-existence region in the
experimental phase diagram, it is not possible to pin down the
critical fields more precisely. We can still make some estimates
based on the above analytical formulas of the critical fields. The
observed spiral wave length of 90nm and the unit cell size of
$a\sim$4.5\AA ~ gives the ratio $D/J = 2\pi (a/\lambda) \approx
1/30$. The critical fields are of order $D^2/J = J (D/J)^2 \sim
J/900$. If we take the observed paramagnetic transition temperature
of $\sim$30K as a measure of $J$, we would get $B_c \sim$ 30K/900
$\sim$ (1/30)K which in magnetic field unit becomes $\sim$ (1/30)T,
in excellent agreement with the observe field ranges of 40-80 mT for
the Skyrmion lattice. The criteria derived in Eq. (\ref{eq:Bc1-Bc2})
is general, applicable to a wide range of chiral magnets
characterized by both $J$ and $D$ without the strong spin anisotropy
effects. For such systems the Skyrme crystal phase formation is
expected in the field range of $D^2/J$.

\begin{figure}[tps]
\includegraphics[scale=0.4]{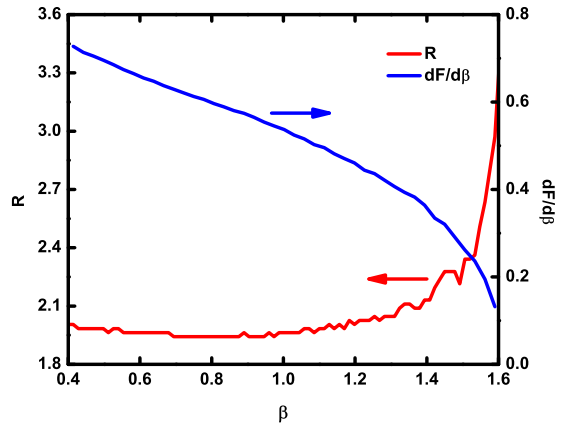}
\caption{(color online) The inter-Skyrmion distance in SkX (in red),
and the first order derivative of SkX's free energy with respect to
the magnetic field $\beta$ (in blue). The distance increases rapidly
near the transition from SkX to FM phase. As the derivative is
nonvanishing at the critical field $\beta_{c2}=1.6$, this transition
is first order.} \label{fig:derivative_and_radius}
\end{figure}

The phase diagram sheds light on the properties of phase transition
as well. Two lines of spiral and SkX have different slopes at the
critical field $\beta_{c1}$, as shown in Fig. \ref{fig:phase}.
Therefore the phase transition between spiral and SkX phase is first
order due to the discontinuity of $\partial F/\partial\beta$ across
the phase boundary. For the transition from SkX to FM, we carefully
examined the derivative $\partial F/\partial\beta$ of the SkX phase
as shown in Fig. \ref{fig:derivative_and_radius}. At the critical
field $\beta_{c2}=1.6$, $\partial F/\partial\beta$ of SkX is clearly
nonvanishing. However the free energy of the FM phase is set to be
zero already, so $\partial F/\partial\beta=0$ on the FM side.
Consequently the transition between SkX and ferromagnetic phases is
first order as well. This is quite consistent with the experimental
results\cite{tokura}. Another interesting issue is the deformation of
Skyrmions at the phase transition from SkX to FM. Half of the
optimized inter-Skyrmion distance, denoted by $R$, in SkX is shown in
Fig. \ref{fig:derivative_and_radius}. Deep inside the SkX phase, the
inter-Skyrmion distance is roughly unchanged, which is just the
optimal distance $2R_0$ controlled by the Dzyaloshinskii-Moriya
physics as discussed above. However when the magnetic field
approaches the critical value $\beta_{c2}$, the inter-Skyrmion
distance increases very rapidly. In this situation, the configuration
of the individual Skyrmion resembles the one shown in the right inset
of Fig. \ref{fig:density}. The ferromagnetic tail grows significantly
while leaving the core region almost unchanged around radius $R_0$.
The growing FM tail near $\beta_{c2}$ indicates that the Skyrmions
are separated far away by the intervening FM phase, and become more
and more dilute when the magnetic field approaches $\beta_{c2}$.

One should be careful to distinguish the SkX with $R\gg R_0$ from the
FM state. In both states the average magnetization is fully
saturated, but there is a qualitative difference in the two phases
because of the existence of topological defects in the SkX. This
explains why the SkX$\rightarrow$FM transition is still first-order
(one cannot turn off topological defects smoothly), despite the fact
that the magnetization reaches unity in a continuous manner at the
SkX/FM phase boundary. A careful examination of the Skyrmion size and
the inter-Skyrmion distance in the thin-film chiral magnet near the
upper critical field is expected to confirm the field dependence of
the two length scales $R_0$ and $R$ discussed in this subsection.
%When $\beta>\beta_{c2}$, the inter-Skyrmion
%distance becomes infinitely long, so that FM phase is dominant.
%Actually Skyrmions are unstable in this situation, and can be
%destroyed by small perturbations, leaving FM the only true ground
%state. This deformation of SkX during the transition to FM may be
%traced experimentally.

\subsection{Analogy to Abrikosov Lattice}
\label{subsection-b}

In Sec. \ref{sec:CP1}, we briefly pointed out the
vortex$\leftrightarrow$Skyrmion correspondence established by the
CP$^1$ mapping (See Fig. \ref{fig:skyrmion-vortex-map}). In this
subsection, we argue that the correspondence in fact extends to the
case of their respective lattice structures.

Abrikosov put forward the lattice solution of vortices in a type-II
superconductor of the form\cite{abrikosov}

\ba \psi (x,y) = \sum_{j=-\infty}^{\infty} c_j e^{2\pi i  (j y/l_y)}
e^{-(x-j l_x)^2 /2\xi^2}, \label{eq:Abrikosov} \ea
with some constants $c_j$. Here $\xi$ is the correlation length,
$l_x$, $l_y$ are the inter-vortex separations in the $x$- and
$y$-directions, respectively, with the relation $l_x l_y = h/eB =
2\pi l_B^2$, $l_B$=magnetic length. We now pose the question: can
one construct a spinor solution $\v z_\mathrm{SkX} (x,y)$ in the
same spirit as Abrikosov's vortex lattice solution, such that the
associated spin configuration $\v n_\mathrm{SkX} =\v
z_\mathrm{SkX}^\dag \bm \sigma \v z_\mathrm{SkX}$ is the Skyrmion
lattice?

We start by writing down the saddle-point equation derived by
minimizing the total energy, $E$. The constraint $\v z^\dag \v z =
1$ is implemented by augmenting the total energy with the Lagrange
multiplier field $\lambda(\v r)$: $E \rightarrow H = E + \int d^2 \v
r ~ \lambda (\v z^\dag \v z - 1)$. Taking the variational derivative
$\delta H / \delta \v z^\dag = 0$ yields the saddle-point equation

\ba 2J \Bigl(\bm \nabla \!-\! i \v A \!-\! i\kappa \bm \sigma
\Bigr)^2 \v z \!+\! 2i D (\v n \cdot \bm \nabla ) \v z \!+ \! (\v B
\cdot \bm \sigma) \v z \!=\! \lambda \v z . \label{eq:simplified}\ea
As a self-consistency check, one can show that the helical spin
solution (\ref{eq:SS-answer}) indeed satisfies the equation when $\v
B = 0$.

The Skyrme crystal state generates a non-zero Skyrmion number that
translates into a non-zero effective magnetic field, $\partial_x A_y
- \partial_y A_x \neq 0$, which in turn should generate a
Landau-level-type solution. The actual equation is non-linear,
however, and the underlying Landau level structure is not clear. The
non-linearity arises from two sources. One is that the vector
potential $\v A$ depends on the knowledge of the solution $\v z$
itself. The second is the presence of $2i D (\v n \cdot \bm \nabla )
\v z$ in the equation. In regard to the second issue we are reminded
the fact that the SkX state exists under a finite magnetic field
which partially polarize the spins. Numerical calculation shows that
the mean moment in the SkX state can be a good fraction of the full
moment\cite{YONH}. It thus appears reasonable to take a spatial
average of $\v n$ and obtain $\v n \cdot \bm \nabla \v z \rightarrow
\langle \v n \rangle \cdot \bm \nabla \v z = 0$, since the spatial
gradient in the two-dimensional lattice is orthogonal to the average
moment direction along $\hat{z}$. Such a conclusion will not
generally hold for 3D lattice or for field tilted away from the
$\hat{z}$-direction in the 2D lattice.

Provided the important character of the solution is not lost upon
the removal of $2i D (\v n \cdot \bm \nabla ) \v z$, one can solve
instead of Eq. (\ref{eq:simplified}) the following problem

\ba 2J \Bigl(\bm \nabla \!-\!i \v A \!-\! i\kappa \bm \sigma \Bigr)^2
\v z \!+\! B \sigma_z  \v z \!= \!\lambda\v z .\ea
We further assume that the fictitious field produced by $\v A$ is
uniform, and choose the Landau gauge $A_x = 0$, $A_y = -Hx$. The
problem of self-consistently deciding the vector potential $\v A$ is
reduced to that of a single constant $H$, answering to the first
source of non-linearity pointed out earlier. As in the typical Landau
level problem we introduce a plane-wave solution for the
$y$-component, $\v z (x,y) = e^{ik y} \v z (x) $, where $\v z(x)$
obeys

\ba 2J \Bigl(\partial_x \!-\! i\kappa \sigma_x \Bigr)^2 \v z \!-\!
2J \Bigl(k \!+\! H x \!-\! \kappa \sigma_y \Bigr)^2 \v z  \!+\! B
\sigma_z \v z \!=\! \lambda\v z . \label{eq:1D-eq}\ea
We have verified that $H>0$ corresponds to the (over)screening of the
external field $B$ by the induced field $H$.

Introducing the  magnetic length $l_H = 1/\sqrt{H}$, and $x_k = x + k
l_H^2$, one can derive the solution of Eq. (\ref{eq:1D-eq}) in the
form $z_1 = \phi_n (x_k /l_H) $, $z_2 = i d_n \phi_{n+1} (x_k /l_H) $
where $\phi_n$ is the $n$-th oscillator wave function and $d_n$ is a
coefficient

\ba  d_n = {2\kappa\sqrt{2H (n\!+\!1)} \over  H\!+\!B/2J \!+\!
\sqrt{( H \!+\! B/2J)^2 \!+\! 8 (n\!+\!1) \kappa^2 H }}.
\label{eq:lambda-and-d} \ea
Focusing on the lowest Landau level solution, $n=0$, the
single-particle wave function obtained reads

\ba \v z (x_k, y) = e^{ik y} \bpm  \phi_0 (x_k /l_H ) \\ i d_0 ~
\phi_1 (x_k /l_H ) \epm .\ea
Wave functions with different $k$'s are degenerate, and can be
grouped into a linear combination in the manner of Abrikosov
solution, Eq. (\ref{eq:Abrikosov}):

\ba  && \v z_\mathrm{SkX} \!=\! \Bigl({2l_x\over(1\!+\! d_0^2)l_H
\sqrt{\pi} }\Bigr)^{1/2} \times\nn
&&  \sum_{j=-\infty}^\infty c_j e^{i2\pi j y/l_y } \bpm
e^{-x_j^2/(2l_H^2)} \\ i d_0\sqrt{2} x_j e^{-x_j^2/(2l_H^2)}/l_H
\epm \! . \label{eq:z-lattice-solution}\ea
Here $k$ is quantized as $k= 2\pi j /l_y$, $j$=integer, and $x_j$
abbreviates $x\!+\! j l_x$. Comparison with the vortex lattice
solution shows that $l_H$ serves as the correlation length $\xi$, as
well as the Skyrmion lattice spacing through the condition $l_x l_y
= 2\pi l_H^2$. The overall constant reflects the average
normalization $(1/\mathrm{Area}) \int dx dy ~ \v z^\dag \v z =
\langle \v z^\dag \v z \rangle = 1$.

We will now show that the state written down in Eq.
(\ref{eq:z-lattice-solution}) captures all essential aspects of the
Skyrme crystal state. Rather than trying to justify the various
approximations that led to Eq. (\ref{eq:z-lattice-solution}), we
regard it as a variational state, whose energy can be checked
against those of other possible spin states. In this way we can
construct a phase diagram similar to the one shown in the previous
subsection.

For the triangular array of Skyrmions, $c_j$ is chosen equal to $1$
and $i$ for even and odd integers, respectively, and $l_y$ equal to
$\sqrt{3}\ l_x/2$. Energy per area ${\cal E}$ evaluated by inserting
the variational solution, Eq. (\ref{eq:z-lattice-solution}), into
Eq. (\ref{eq:energy-density-in-CP1}) and carrying out the spatial
integration reads

\ba {\cal E}_\mathrm{SkX} =2J \left( 2\kappa^2\!-\!{4\sqrt{2H}\kappa
d_0\over 1+d_0^2}\!+\!{1+3d_0^2\over 1+d_0^2}H \right)
\!-\!B{1-d_0^2\over 1+d_0^2}.\nn \label{eq:SkX-energy} \ea
The extremum condition $\partial {\cal E}_\mathrm{SkX} /\partial H =
0$ uniquely fixes $H$, hence the magnetic length $l_H$. One can read
off the relevant energy scales better by dividing out both sides by
$4J \kappa^2 = D^2/J$:

\ba {{\cal E}_\mathrm{SkX} \over D^2 /J}  &=&1 \!-\!{2\sqrt{2}
d_0\over 1+d_0^2}{1\over \kappa l_H } \!+\!{1+3d_0^2\over
1+d_0^2}{1\over 2\kappa^2 l_H^2  } \!-\! b {1-d_0^2\over 1+d_0^2},\nn
d_0 &=& {2\sqrt{2}\kappa l_H \over 1\!+\! 2 \kappa^2 l_H^2 b \!+\!
\sqrt{(1 \!+\! 2 \kappa^2 l_H^2 b )^2 \!+\! 8 \kappa^2 l_H^2 }}.
\label{eq:scaled-SkX-energy} \ea
We used the field in reduced unit $b=B/(D^2/J)$ in the above
expressions. The re-scaling makes it clear that the relevant Zeeman
energy is indeed $D^2/J$, in agreement with the analysis of the
previous subsection.

\begin{figure}[ht]
\includegraphics[scale=0.4]{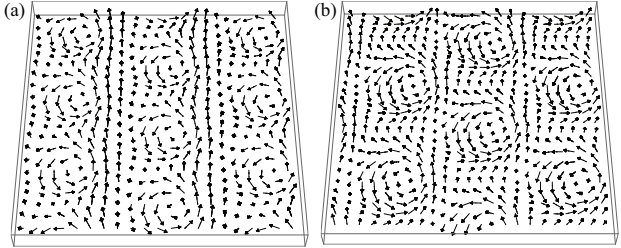}
\caption{(color online) (a) A typical Skyrme crystal spin
configuration given by Eq. (\ref{eq:z-lattice-solution}) with
$B=D^2/J$, with the optimized lattice spacing $\kappa l_H
=\sqrt{3/2}$. (b) Skyrme crystal spin configuration obtained by
Monte Carlo method from the lattice spin model\cite{YONH}. }
\label{fig:CP1-skyrmion}
\end{figure}

Figure \ref{fig:CP1-skyrmion} (a) shows the typical spin
configuration given out by the CP$^1$ solution $\v z_\mathrm{SkX}$
for $b=1$ $(B=D^2/J)$. Optimizing the energy gives out $\kappa l_H =
\sqrt{3/2}$. Spins are pointing up, aligned with the $B$ direction,
away from the Skyrmion center and pointing down at the core. The
sense of spin swirling (vorticity) in the core region is consistent
with the right-handed proper screw direction of the helical spin
phase. Reversing the sign of $\kappa$, hence $d_0 \rightarrow -d_0$
in Eq. (\ref{eq:z-lattice-solution}), leads to the left-handed screw
and a clockwise swirling of spins near the cores. For comparison a
typical Skyrme crystal configuration produced from the Monte Carlo
annealing of the lattice spin model\cite{YONH} is reproduced in Fig.
\ref{fig:CP1-skyrmion} (b). It is clear that the essential features
of the Skyrmion lattice configuration has survived the several
drastic approximations employed in arriving at the Skyrme crystal
solution $\v z_\mathrm{SkX}$.

\begin{figure}[h]
\includegraphics[scale=1.0]{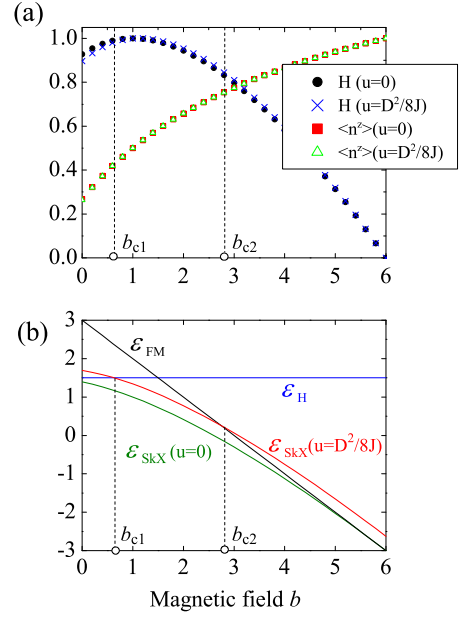}
\caption{(color online) (a) Self-consistently induced field $H$ (in
the dimensionless unit $\kappa^2 H$) and the ferromagnetic
polarization $\langle n^z \rangle$, against external magnetic field
$b = B/(D^2/J)$. Taking $\kappa l_H =\sqrt{3/2}$, two different
values of $u$ (see text for definition), $u=0$ and $u=(1/8)(D^2/J)$
were used with little differences in the results. (b) Energies of
helical, Skyrme crystal, and ferromagnetic spin states against $b$,
both measured in units of $D^2/J$. SkX energy $\cal E_\mathrm{SkX}$
goes up with $u$ while the other energies remain insensitive to $u$.
$b_{c1}$ and $b_{c2}$ define the two first-order transitions. }
\label{fig:energy-vs-B}
\end{figure}

Self-consistently determined $H$ and the average polarization
$\langle n^z \rangle = \langle \v z^\dag \sigma^z \v z\rangle =
(1-d_0^2)/(1+d_0^2)$ against the external field $b$ are shown in
Fig. \ref{fig:energy-vs-B} (a). It is seen that $H$ tends to zero as
the external field drives full polarization of spins, $\langle n_z
\rangle \approx 1$. The core size and the Skyrmion spacing, both of
which are fixed by $l_H$, diverges accordingly. In reality the
divergence is cut off by a first-order transition to the
energetically more favorable FM state at the upper critical field
$B_{c2}$.

Shown in Fig. \ref{fig:energy-vs-B} (b) are the energies of the
three main competing spin configurations - helical, Skyrme crystal,
and ferromagnetic spin states - plotted against $b$. The SkX state
gives the lowest energy regardless of the field strength, in
contrast to both Monte Carlo\cite{YONH} and
experimental\cite{tokura} findings showing the ground state
evolution helical$\rightarrow$SkX$\rightarrow$FM with increasing
$B$. We also find that the energy is insensitive to the crystal
structure of the Skyrmions being a square or a triangular lattice.
%\ba u {2+4d_0^2-d_0^4\over 2(1+d_0^2)^2} + \cdots . \ea
%

Both these problems can be remedied by re-visiting the constraint
$\v z^\dag  \v z =1$, which is so far imposed at the crudest,
mean-field level. Helical and FM spin solutions obey the constraint
exactly anyway, but the SkX solution does not. An inclusion of the
potential term $u(\v z^\dag \v z -1)^2$ ($u>0$) in the free energy
density (\ref{eq:energy-density-in-CP1}), which imposes the
constraint at the local level, would increase the energy density of
the SkX solution by $u [ \langle (\v z^\dag_\mathrm{SkX} \v
z_\mathrm{SkX} )^2 \rangle \!-\! 1 ]$. As seen in Fig.
\ref{fig:energy-vs-B} (b), a suitable choice of $u$ restores the
correct sequence of ground states punctuated by two critical fields,
$B_{c1}$ for helical$\rightarrow$SkX and $B_{c2}$ for
SkX$\rightarrow$FM transitions. According to Fig.
\ref{fig:energy-vs-B} (a), $H$ remains close to unity in the whole
SkX region, implying the inter-Skyrmionic spacing $l_H$ comparable
to the period of the spin spiral $\lambda \sim \kappa^{-1}$. This is
indeed the case both in the Monte Carlo calculation\cite{YONH} and
experimentally\cite{hexagonal-SC,tokura}. The chance of a larger
Skyrmion spacing as $H$ becomes very small is pre-emptied by the
transition to a ferromagnetic phase with a lower energy at
$B=B_{c2}$. Finally, the average $\langle (\v z^\dag_\mathrm{SkX} \v
z_\mathrm{SkX} )^2 \rangle$ calculated at a number of $(B,H,\kappa)$
values is in favor of the triangular lattice having a lower energy
over the square lattice. This is the same criterion used by
Abrikosov\cite{abrikosov} and subsequent workers\cite{kleiner} in
identifying the most stable lattice structure of vortices by
calculating the average of $|\psi|^4$ in the U(1) GL theory. Unlike
in Abrikosov's case, the average $\langle (\v z^\dag_\mathrm{SkX} \v
z_\mathrm{SkX} )^2 \rangle$ depends on the $(B,H,\kappa)$ values
used and are not universal.
%0.936556 and 0.841026 ($l_H=1,~\kappa=\sqrt{3/2}$), making the
%triangular lattice configuration slightly lower in energy.

\begin{figure}[h]
\includegraphics[scale=0.4]{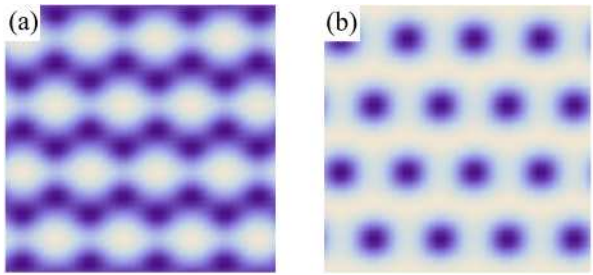}
\caption{(color online) (a) Induced field distribution
$H_\mathrm{SkX}(x,y)$ from the Skyrmion lattice solution $\v
z_\mathrm{SkX}$. Bright area is the Skyrmion core where the
intensity is the maximum. (b) Distribution of $n^z (x,y)=\v
z^\dag_\mathrm{SkX} \sigma^z \v z_\mathrm{SkX}$. Dark area means
reversed spins at the Skyrmion core. } \label{fig:H-SkX}
\end{figure}

With $\v z_\mathrm{SkX}$ at hand, the induced field distribution

\ba H_\mathrm{SkX}(x,y) &= & i (\partial_x \v z^\dag_\mathrm{SkX}
\partial_y \v z_\mathrm{SkX} - \partial_y \v z^\dag_\mathrm{SkX}
\partial_x \v z_\mathrm{SkX} )\nn
&& - (2\pi/ l_y )( x / l_x )
\partial_x (\v z^\dag_\mathrm{SkX}\v z_\mathrm{SkX} ) \ea
can be worked out\cite{comment2}. Figure \ref{fig:H-SkX} (a) gives
the field distribution, along with the distribution of the
$z$-component of the local magnetization $n^z (x,y)$ in Fig.
\ref{fig:H-SkX} (b). The field intensity reaches a maximum at the
Skyrmion core and nearly equals zero in the FM background. The
spatial average of $H_\mathrm{SkX}(x,y)$ is approximately equal to
$H$ used as an input in $\v z_\mathrm{SkX}$.

We should emphasize that in the 3D chiral magnets it is the conical
phase, with the propagation vector along the $\hat{z}$-direction,
having the lower energy and hence occupying the lower temperature
side of the phase diagram instead of the SkX state, under moderate
magnetic field\cite{hexagonal-SC,FeCoSi,B20-general}. The conical
phase cannot exist in a 2D system, and this probably contributes to
the SkX state being realized at low temperature for thin-film
samples. It still remains an interesting challenge how one can
understand the crossover in the behavior from 2D to 3D systems.

\subsection{Comparison to earlier work}
Bogdanov and collaborators have extensively investigated the
possibility of the Skyrmion formation in chiral magnets on the basis
of Ginzbug-Landau models with Dzyaloshinskii-Moriya
interaction\cite{bogdanov1,bogdanov2}. In particular the field
dependence of the phase diagram presented in Ref.
\onlinecite{bogdanov2} (Fig. 9 in their paper) correctly captures the
spiral$\rightarrow$SkX$\rightarrow$FM phase change recently
investigated by Monte Carlo method\cite{YONH},
experimentally\cite{tokura}, and in the previous two subsections. The
analysis of subsection \ref{subsection-a} amounts to their
``circular-cell" approximation.

In Ref. \onlinecite{bogdanov2} the nature of the phase transition
from SkX to FM was not determined conclusively. Based on our analysis
presented in Secs. \ref{subsection-a} and \ref{subsection-b}, it is
clearly first-order. Physically this should be clear since there is
no way to turn off topological defects in a continuous fashion. Nor
is it possible to fuse a Skyrmion with an anti-Skyrmion to annihilate
them, since the system under consideration consists only of one
species of
Skyrmions\cite{hexagonal-SC,FeCoSi,B20-general,YONH,tokura}.
Experiments also support the first-order phase boundary\cite{tokura}.

In their analysis, Bogdanov \textit{et al.} assumed a
two-dimensional structure homogeneously extended along the third
direction. Our model is explicitly two-dimensional. With the recent
experimental input\cite{hexagonal-SC} we know that for known 3D
chiral magnets such as MnSi, in the field range where the SkX phase
(a.k.a. A-phase) is found, the low-temperature state has the conical
spin structure. In order to stabilize the Skyrmionic phase it is
therefore essential to suppress the three-dimensionality as well as
to turn on a magnetic field. The present investigation fills in the
gap that existed between the early theoretical work and the new
insights offered by recent experimental progress.

Finally, our paper presents a new framework for dealing with chiral
magnetic systems. The CP$^1$ formulation allows us to understand the
Skyrmion lattice formation in direct analogy to the Abrikosov vortex
problem in superconductors, and to the Landau-level physics. It is
our belief that the new formulation can be used in conjunction with
the conventional classical vector theory in studying other aspects
of chiral magnetism such as their dynamics.

\section{Discussion}
\label{sec:discussion}

Although the detailed examination of the thermal effects at nonzero
temperature is outside the scope of this paper, we can make
qualitative assessment about the thermal fluctuations for the Skyrme
crystal state. Both translational and spin rotational symmetries are
broken in the SkX state, and one accordingly expects two types of
Goldstone modes: phonon-like displacements of Skyrmion positions and
spin waves.

Applying the Lindemann criterion, the melting of the Skyrmion
lattice will take place if the thermal fluctuation in the
center-of-mass position of a Skyrmion becomes comparable to the
inter-Skyrmion spacing. As discussed in the previous section, the
typical magnetic energy difference per spin for the three magnetic
phases - helical, SkX, and FM - are of the order $D^2/J$. A shift of
the Skyrmion position by one lattice unit increases the energy per
spin by a typical amount $D^2 /J$, or $J l_H^{-2}$ if the linear
size of the Skyrmion $l_H \sim \kappa^{-1} \sim J/D$ is used. The
energy increase associated with the Skyrmion displacement by $x$
should then be $J \times (x/l_H )^2$. Applying the equipartition
theorem we obtain $J \times \langle x^2 \rangle /l_H^2 \sim T$ at a
temperature $T$, and the melting temperature $T_m \sim J$ from the
condition $\sqrt{\langle x^2 \rangle } \sim l_H$.

Another possible disordering mechanism is the spin fluctuation. In
the ground state each Skyrmion unit cell exhibits identical spin
configuration. If the spin orientation of one particular Skyrmion
unit cell is completely reversed, the cost in Zeeman energy is
roughly $B$ times the number of spins, or $B l_H^2$. Noting that $B$
in the relevant SkX phase is $D^2 /J$, we conclude that the
single-Skyrmion spin flip has the energetic cost of $J$. From this
we conclude that the coupling energy of adjacent Skyrmions must be
$J \times (\Delta \theta /\pi)^2$ for a small angle difference
$\Delta \theta$ of the two nearby Skyrmion spin orientations. By
invoking equipartition theorem again, we arrive at a spin-melting
temperature $T_m \sim J$. In conclusion, destroying the global
ordering of the SkX state by either center-of-mass displacement or
spin fluctuation requires a temperature of order $J$, even though
the magnetic field energy difference between the various phases is
only $D^2/J$. It is the large size of the Skyrmion, of order
$(J/D)^2$, which compensates for the small magnetic energy scale and
renders a sizable melting temperature.

To summarize, we have provided a theoretical treatment of the Skyrme
crystal phase recently observed in the thin film of chiral magnet
Fe$_{0.5}$Co$_{0.5}$Si\cite{tokura}. Two independent constructions
of the Skyrmion lattice state were used to calculate the energy of
this state and compare it against those of other competing states,
namely helical spin and ferromagnetic spin, as the perpendicular
magnetic field strength is increased. The resulting phase diagram
giving the successive ground state evolution of
helical$\rightarrow$SkX$\rightarrow$FM phases is in accord with the
Monte Carlo simulation\cite{YONH} and the thin-film
experiment\cite{tokura}. We gave a general argument for the two
critical field strengths separating the SkX state from the helical
state on the low-field side and from the FM state on the high-field
side to be of order $D^2/J$. Other spiral magnets with similar spin
structures are expected to exhibit Skyrme crystal phase for field
ranges of $D^2/J$ if they are made in the thin-film form. We also
gave a qualitative argument why the melting of the Skyrme crystal
should occur at $T_m \sim J$ despite the small magnetic energy scale
$D^2/J$ needed to stabilize the SkX phase. As an interesting
by-product, we derived the variational wave function corresponding
to the Skyrme crystal state in the CP$^1$ form.

\acknowledgments  N. N. is supported by Grant-in-Aids for Scientific
Research (No. 17105002, 19019004, 19048008, 19048015, and 21244053)
from the Ministry of Education, Culture, Sports, Science and
Technology of Japan, and also by Funding Program for World-Leading
Innovative R\&D on Science and Technology (FIRST Program). H. J. H.
is supported by Mid-career Researcher Program through NRF grant
funded by the MEST (No. R01-2008-000-20586-0), and in part by the
Asia Pacific Center for Theoretical Physics.

%H. J. H. is supported by the Korea Research Foundation Grant
%(KRF-2008-521-C00085, KRF-2008-314-C00101) and in part by the Asia
%Pacific Center for Theoretical Physics.


\begin{thebibliography}{99}
%

\bibitem{skyrme} T. H. R. Skyrme, Proc. Roy. Soc.
(London) A \textbf{260}, 127 (1961); Nuc. Phys. \textbf{31}, 556
(1962).


\bibitem{sondhi} S. L. Sondhi, A. Karlhede, S. A. Kivelson, and E.
H. Rezayi, Phys. Rev. B \textbf{47}, 16419 (1993).

\bibitem{skyrmion-evidence} S. E. Barrett, G. Dabbagh, L. N. Pfeiffer, K. W. West,
and R. Tycko, Phys. Rev. Lett. \textbf{74}, 5112 (1995); E. H. Aifer,
B. B. Goldberg, and D. A. Broido, Phys. Rev. Lett. \textbf{76}, 680
(1996); V. F. Mitrovi\'{c}, M. Horvati\'{c}, C. Berthier, S. A. Lyon,
and M. Shayegan, Phys. Rev. B \textbf{76}, 115335 (2007).
%A. Schmeller, J. P. Eisenstein, L. N. Pfeiffer, and K. W. West, Phys.
%Rev. Lett. \textbf{75}, 4290 (1995);

\bibitem{macdonald} L. Brey, H. A. Fertig, R. C\^{o}t\'{e}, and A. H.
MacDonald, Phys. Rev. Lett. \textbf{75}, 2562 (1995).




\bibitem{skyrme-crystal-evidence} G. Gervais, H. L. Stormer, D. C. Tsui, P. L. Kuhns,  W. G.
Moulton, A. P. Reyes, L. N. Pfeiffer, K. W. Baldwin, and K. W. West,
Phys. Rev. Lett. \textbf{94}, 196803 (2005); Yann Gallais, Jun Yan,
Aron Pinczuk, Loren N. Pfeiffer, and Ken W. West, Phys. Rev. Lett.
\textbf{100}, 086806 (2008); Han Zhu, G. Sambandamurthy, Yong P.
Chen, P. Jiang, L. W. Engel, D. C. Tsui, L. N. Pfeiffer, and K. W.
West, Phys. Rev. Lett. \textbf{104}, 226801 (2010).


%For a review of the subject, see \textit{Perspective in Quantum Hall
%Effects}, edited by Sankar das Sarma and Aron Pinczuk (John Wiley \&
%Sons, 1997).

\bibitem{hexagonal-SC} S. M\"{u}hlbauer, B. Binz, F. Joinetz, C.
Pfleiderer, A. Rosch, A. Neubauer, R. Georgii, and P. B\"{o}ni,
Science \textbf{323}, 915 (2009).

\bibitem{FeCoSi} W. Munzer, A. Neubauer, T. Adams,
S. Muhlbauer, C. Franz, F. Jonietz, R. Georgii, P. Boni, B.
Pedersen, M. Schmidt, A. Rosch, and C. Pfleiderer, Phys. Rev. B
\textbf{81}, 041203(R) (2010).

\bibitem{B20-general} C. Pfleiderer, T. Adams, A. Bauer, W.
Biberacher, B. Binz, F. Birkelbach, P. B\"{o}ni, C. Franz, R.
Georgii, M. Janoschek, F. Jonietz, T. Keller, R. Ritz, S.
M\"{u}hlbauer, W. M\"{u}nzer, A. Neubauer, B. Pedersen, and A.
Rosch, J. Phys.:Condens. Matter \textbf{22}, 164207 (2010).


\bibitem{AHE} A. Neubauer, C. Pfleiderer, B. Binz, A. Rosch, R.
Ritz, P. G. Niklowitz, and P. B\"{o}ni, Phys. Rev. Lett.
\textbf{102}, 186602 (2009).

\bibitem{YONH} Su Do Yi, Shigeki Onoda, Naoto Nagaosa, and Jung Hoon Han, Phys. Rev. B
\textbf{80}, 054416 (2009).

\bibitem{tokura} X. Z. Yu, Y. Onose, N. Kanazawa, J. H. Park, J. H.
Han, Y. Matsui, N. Nagaosa, and Y. Tokura, Nature (London)
\textbf{465}, 901 (2010).


\bibitem{bogdanov1} A. N. Bogdanov and D. A. Yablonskii, Sov. Phys. JETP
\textbf{68}, 101 (1989).

\bibitem{bogdanov2} A. Bogdanov and A. Hubert, J. Magn. Magn. Mater. \textbf{138}, 255
(1994).

\bibitem{bak} Per Bak and M. H{\o}gh Jensen, J. Phys. C \textbf{13},
L881 (1980).

\bibitem{abrikosov} A. A. Abrikosov, Sov. Phys. JETP \textbf{5}, 1174 (1957).

\bibitem{tonomura} A. Tonomura, H. Kasai, O. Kamimura, T. Matsuda, K. Harada,
Y. Nakayama, J. Shimoyama, K. Kishio, T. Hanaguri, K. Kitazawa, M.
Sasase and S. Okayasu, Nature \textbf{412}, 620 (2001).

\bibitem{auerbach}
%Assa Auerbach, \textit{Interacting Electrons and
%Quantum Magnetism} (Springer-Verlag, Berlin, 1994).
N. Nagaosa, \textit{Quantum Field Theory in Strongly Correlated
Electronic Systems}, Chap. 5 (Springer, 1999).


\bibitem{comment1} It is instructive to note that the original
analysis of Skyrme (Ref. \onlinecite{skyrme}) was carried out in the
language which amounts to what we call the CP$^1$ formulation.


\bibitem{Raj} R. Rajaraman, \textit{Solitons and Instasntons}, Chap. 3 (North Holland, 1987).


%\bibitem{boni} A. Neubauer, C. Pfleiderer, B. Binz, A. Rosch, R.
%Ritz, P. G. Niklowitz, and P. B\"{o}ni, Phys. Rev. Lett.
%\textbf{102}, 186602 (2009).


\bibitem{kleiner} W. H. Kleiner, L. M. Roth, and S. H. Autler, Phys.
Rev. \textbf{133}, A1226 (1964).


\bibitem{comment2} The $(2\pi/l_y ) (x/l_x) \partial_x (\v
z^\dag_\mathrm{SkX}\v z_\mathrm{SkX} )$ restores the lattice
translational invariance of $H_\mathrm{SkX}(x,y)$ and would be
unnecessary if the constraint $\v z^\dag \v z=1$ was obeyed exactly.



\end{thebibliography}
\end{document}